\documentclass[a4paper,11pt]{article}
\usepackage{jinstpub} % for details on the use of the package, please see the JINST-author-manual
\usepackage{lineno}
\usepackage[export]{adjustbox}
\usepackage[section]{placeins}
\usepackage{comment}
\usepackage{orcidlink}

\title{High-bandwidth frequency domain multiplexed readout of transition-edge sensors for neutrinoless double beta decay searches}

 \author[a,b,*]{M. Adamič\orcidlink{0009-0004-5822-088X},\note[*]{Corresponding author.}}
 \author[c,1]{M. Beretta\orcidlink{0000-0002-9324-1959},\note{Now at INFN Sezione di Milano-Bicocca, Milano, Italy.}}
 \author[b,d,*]{J. Camilleri\orcidlink{0000-0002-0648-4291},}
 \author[b,2,*]{C. Capelli\orcidlink{0000-0003-3330-621X},\note{Now at Physik-Institut, University of Zürich, 8057 Zürich, Switzerland.}}
 \author[a]{M. A. Dobbs\orcidlink{0000-0001-7166-6422},}
 \author[e]{T. Elleflot\orcidlink{0000-0002-5166-5614},}
 \author[b]{B. K. Fujikawa\orcidlink{0000-0002-7001-717X},}
 \author[b,c]{Yu. G. Kolomensky\orcidlink{0000-0001-8496-9975},}
 \author[f]{D. Mayer\orcidlink{0000-0001-6149-4196},}
 \author[a]{J. Montgomery\orcidlink{0000-0003-0345-8386},}
 \author[g]{V. Novosad\orcidlink{0000-0003-2127-2374},}
 \author[c,h]{A. M. Sindhwad\orcidlink{0009-0008-5090-9796},}
 \author[c]{V. Singh\orcidlink{0000-0002-0401-4421},}
 \author[i]{G. Smecher\orcidlink{0000-0002-5560-187X},}
 \author[e]{A. Suzuki\orcidlink{0000-0001-8101-468X},}
 \author[c]{B. Welliver\orcidlink{0000-0002-4459-4563}}

 \affiliation[a]{Department of Physics and Trottier Space Institute, McGill University, Montreal, QC H3A 2T8, Canada}
 \affiliation[b]{Nuclear Science Division, Lawrence Berkeley National Laboratory, Berkeley, CA 94720, USA}
 \affiliation[c]{Department of Physics, University of California Berkeley, Berkeley, CA 94720, USA}
 \affiliation[d]{Center for Neutrino Physics, Virginia Polytechnic Institute and State University, Blacksburg, VA 24061, USA}
 \affiliation[e]{Physics Division, Lawrence Berkeley National Laboratory, Berkeley, CA 94720, USA}
 \affiliation[f]{Laboratory for Nuclear Science, Massachusetts Institute of Technology, Cambridge, MA 02139, USA}
 \affiliation[g]{Materials Science Division, Argonne National Laboratory, Lemont, IL 60439, USA}
 \affiliation[h]{Department of Nuclear Engineering, University of California Berkeley, Berkeley, CA 94720, USA}
 \affiliation[i]{t0.technology, Montreal, QC H2J 2L1, Canada}
 
\emailAdd{michel.adamic@mail.mcgill.ca}
\emailAdd{jcamilleri@vt.edu}
\emailAdd{chiara.capelli@physik.uzh.ch}

\abstract{The next-generation of cryogenic neutrinoless double-beta decay experiments require increasingly fast readout in order to improve background discrimination. These experiments, operated as cryogenic calorimeters at $\sim$10\,mK, are usually read out by high-impedance neutron transmutation doped (NTD) thermistors, which provide good energy resolution, but are limited by $\sim$1\,ms response times. Superconducting detectors, such as transition-edge sensors (TESs) with a time resolution of $\sim$100\,$\mu$s, offer superior timing performance over NTD semiconductor bolometers.
To make this technology viable for an application to a thousand or more channels, multiplexed readout is necessary in order to minimize the thermal load and radioactive contamination induced by the readout.
Frequency-domain multiplexing readout (fMUX) for TESs, previously developed at Berkeley Lab and McGill University, is currently in use for mm-wave telescopes with detector sampling rates in the order of 100 Hz.
We demonstrate a new readout system, based on the McGill/Berkeley digital fMux readout, to satisfy the higher bandwidth and noise requirements of the next generation of TES-instrumented cryogenic calorimeters. Each multiplexing readout module comprises 10 superconducting resonators in the 1--5 MHz range and a DC superconducting quantum interference device (DC-SQUID), interfaced to high-speed field programmable gate array (FPGA)-based electronics for digital signal processing and low-latency SQUID feedback. The new readout samples detectors at 156 kHz, three orders of magnitude faster than its cosmology-oriented predecessor, and demonstrates a stable feedback bandwidth of 3 kHz in a real TES-based system.}


\arxivnumber{2601.23106} % Only if you have one

\begin{document}
    \maketitle
\flushbottom

\section{Introduction}
\label{sec:intro}

Cryogenic experiments searching for neutrinoless double beta decay employ increasingly large detector masses and require faster cryogenic sensors to discriminate between the expected signal and the different types of background. Transition edge sensors (TESs), which are prolifically used in cosmic microwave background (CMB) telescopes (e.g. \cite{Dobbs_fmux, POLARBEAR:2015jaw, Thornton:2016wjq, BicepKeck:2021ybl, SPT-3G:2021vps, McCarrick:2021crh}), are a promising technology for both calorimeters and light detectors\,\cite{Singh:2022rck}. Large arrays of TESs have been used in rare event searches for dark matter\,\cite{Pyle:2006rp}. However, when operating at cryogenic temperatures of about 10\,mK, the thermal load from the wiring used to instrument these detectors presents an architectural challenge. Multiplexing the readout becomes necessary with $\mathcal{O}$(1000) detector channels. Additionally, reducing the cabling and material close to the detector is beneficial for radioactive background reduction.

This work discusses the multiplexed readout developed for TES detectors that aim to measure light coming from particle interactions in scintillating crystals. Such an application is particularly relevant for the CUPID (CUORE Upgrade with Particle IDentification) experiment\,\cite{CUPID:2025avs}.
CUPID aims to detect 0$\nu\beta\beta$ from $^{100}$Mo, with the signal appearing as a monoenergetic peak at the Q-value of the reaction (3034\,keV) above a constant background. A large fraction of this background, induced by alpha particles, is discriminated by the CUPID heat/light double readout technology, employing scintillating Li$_2^{100}\mathrm{MoO}_4$ (LMO) crystals coupled to light detectors. However, due to the relatively short half-life of the $2\nu\beta\beta$ decay of $^{100}$Mo, $T_{1/2}^{2\nu} \simeq 7.06 \times 10^{18}$\,yr~\cite{Barabash:2020nck}, the pileup coming from this process would become the largest source of background in the region of interest (ROI) for CUPID, contributing up to half of the total expected background budget\,\cite{CUPID:2025avs}. Consequently, this is one of the most critical items to address to increase the sensitivity of the experiment. 

Timing pileup consists of events that are coincident in a time window shorter than the length of a single pulse. As this background is originating from the same isotope as the process of interest, the only way to reduce it is to discriminate it from single interactions with faster detectors. Although CUPID's baseline design\,\cite{CUPID:2025avs} utilizes light detectors with Neganov-Trofimov-Luke amplification readout through NTD sensors\,\cite{Chernyak:2016aps}, multiplexed TESs are an alternative readout under development, and are expected to be the design choice for the future tonne-scale upgrade of the experiment \,\cite{CUPID_1T}. TESs are one order of magnitude faster than NTDs ($\mathcal{O}(100\,\mu$s) vs $\mathcal{O}$(ms))\,\cite{Singh:2022rck, Chernyak:2016aps} and allow for a significantly better rejection of $2\nu\beta\beta$ pileup events, crucial for tonne-scale $0\nu\beta\beta$ searches.

TES multiplexing is a highly mature and broadly used technology in CMB experiments. There are three main techniques - time division multiplexing (TDM), MHz frequency division multiplexing (FDM or fMUX) and GHz microwave multiplexing ($\mu$MUX). Telescopes from the SPT-3G, POLARBEAR-2/Simons Array and LiteBIRD family use digital fMUX based on the "Iceboard" readout system originally developed by Berkeley and McGill \cite{Bender:2014nnc}. Alternatively, ACT \cite{Thornton:2016wjq} and BICEP/Keck Array \cite{BicepKeck:2021ybl, BICEP:2023jup} employ time-division multiplexing based on University of British Columbia's MCE system, while the newly commissioned Simons Observatory boasts $\mu$MUX readout \cite{McCarrick:2021crh} powered by SLAC's SMuRF electronics. These CMB experiment readout systems demonstrate very high multiplexing factors (in the order of 30-100 for TDM/FDM, and even 1000 for $\mu$MUX), but all measure very slow signals. Therefore, the detector timestreams are downsampled to $\sim$100 Hz, a much too low bandwidth for calorimetric applications.

The development of high-bandwidth multiplexing readout systems has been mostly driven by X-ray and gamma-ray astronomy, where TESs are used in microcalorimeter arrays for spectroscopy. NIST has been working on fast TDM readout of TESs for ESA's Athena mission X-IFU instrument \cite{DurkinAthenaTDM}, reporting a detector frame rate of 156.25 kHz and TES risetimes of 60 $\mu$s \cite{Doriese2016XrayTDM}. SRON has been developing FDM readout for the same mission \cite{vanderKuurfMUXAthena}, first demonstrating a 14x fMUX \cite{Akamatsu2020fMUXAthena} and finally an improved 37x fMUX \cite{Akamatsu:2021vcs} version of the system, sampling detector timestreams at 156.25 kHz. The readout uses Baseband Feedback (BBFB) \cite{denHartogBBFeedbackSRON} to linearize the DC-SQUID, and demonstrates a TES risetime of 180 $\mu$s \cite{Akamatsu:2021vcs}. Lastly, significant improvements in high-bandwidth TES readout have been achieved using microwave multiplexing ($\mu$MUX), for example by NIST \cite{Mates2017uMUX}. Together with INFN, they are developing readout for TES microcalorimeters in the HOLMES neutrino mass experiment \cite{Alpert:2019fit}, demonstrating a 32x MUX system with 500 kHz detector sampling rates and a very impressive recorded TES risetime of 10--20 $\mu$s \cite{Giachero:2021tli}. Similar results have been achieved by groups in Japan in collaboration with SRON, presenting a 38x MUX prototype with 500 kHz sampling and recorded TES risetimes between 6--18 $\mu$s \cite{Nakashima2020uMUXSOA}. Because of very large available total readout bandwidth of several GHz, $\mu$MUX systems offer a significant performance advantage over TDM and FDM systems, with higher MUX factors, higher detector bandwidths and lower crosstalk due to much larger resonance spacing. However, both cold electronics (using RF-SQUIDs) and GHz-range warm electronics are considerably more complex and expensive, and therefore TDM and fMUX systems are still in widespread use.

In this work, we set out to develop a high-bandwidth fMUX readout system for rare-event searches like CUPID, by exploring the possibilities and also the limits of our highly mature Iceboard readout platform \cite{Bandura:2016dpm} that is commonly used in TES-based CMB experiments \cite{Bender:2014nnc, SPT-3G:2021vps, POLARBEAR:2015jaw}. We consider the following requirements for a CUPID-like TES detector readout:
\begin{itemize}
    \item Signal bandwidth in the several kHz range, in order to capture the $\mathcal{O}$(100\,$\mu$s) risetime of TES sensors, and a sampling rate above 100 kHz to allow for pulse shape discrimination techniques. The timing resolution required by CUPID to discriminate pileup such as to reach the required sensitivity is $\sim$170\,$\mu$s. This value is computed considering the counting rate of randomly coincident decay events in a 280\,g LMO crystal at the Q-value of $^{100}$Mo~\cite{Ahmine:2023xhg} and the goal of the pileup component of the desired background index reported in \cite{CUPID:2025avs};
    \item TES detectors require low series impedance readout and cabling to maintain stability of the electro-thermal feedback loop. DC superconducting quantum interference devices (DC-SQUIDs) are therefore used to readout and amplify the signal;
    \item The technology needs to be scalable to tonne-scale calorimeters that will employ thousands of channels;
    \item A multiplexing factor of 10-15 is sufficient for the CUPID cryostat to achieve sufficiently low thermal load at 10\,mK. For the first phase of CUPID, the experiment will use the current CUORE cryostat~\cite{Alduino:2019xia}. In our case the MUX system is placed on the still stage; considering a multiplexing factor of 10, the heat load of the SQUIDs for $\sim$2000 channels is $\mathcal{O}(10)\,\mu$W, well below the cryostat cooling power. However, keeping the multiplexing factor low limits the number of channels that would be lost in case of hardware or wiring failures;
    \item Radioactivity constraints require having as little material as possible close to the detector, meaning that the readout hardware has to be kept as far as possible from the TESs. In the case of the CUORE cryostat, the distance between the SQUIDs on the still stage and the detectors placed below the lead shield on the mixing chamber is on the order of 1-2\,m, depending on where the detector is located within the experimental volume~\cite{Alduino:2019xia}.
\end{itemize}

One of the main challenges compared to CMB telescope implementations using this technology is adapting the design to higher signal bandwidth. In CUPID, the scintillation light signals are orders of magnitude faster than those measured in CMB experiments. This requires a frequency domain multiplexing (fMUX) readout system and thus a resonant circuit with resonance widths on the order of tens of kHz. A second important difference is minimizing material close to the TES to reduce radioactive contamination, which would otherwise mask the rare event signal. Consequently, the cabling between the TES and the SQUID amplification stage needs to be longer, which can shift the detector resonance frequencies and is potentially more susceptible to noise pickup from the environment. In addition, due to the size of the CUPID cryostat, the cabling between the SQUID and the warm readout is much longer than in CMB experiments, which inserts signal attenuation that negatively affects the signal-to-noise ratio. The noise performance of the designed fMUX system is a fairly rich and nontrivial topic, which we discuss in a separate paper -- see\,\cite{Adamic:noise}.

The structure of this paper is as follows. In section~\ref{sec:fmux} we present the general working principle of the developed readout system. Section~\ref{sec:cold} focuses on the cryogenic components, while the warm electronics and the associated field programmable gate array (FPGA) firmware are described in detail in section~\ref{sec:warm}. In section~\ref{sec:measurements} we report the first measurements carried out to validate the system. Section~\ref{sec:conclusion} concludes the paper and provides an outlook on future work.

\section{Frequency domain multiplexing TES readout}
\label{sec:fmux}

\begin{figure}[t]
\centering
\includegraphics[width=\textwidth]{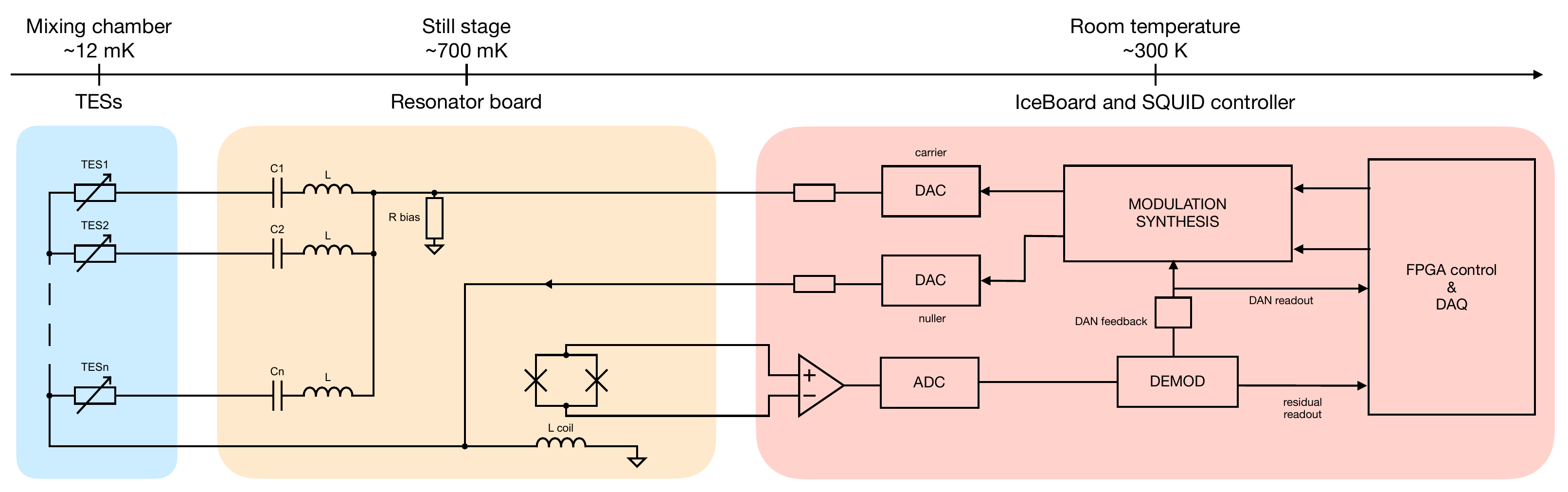}
\caption{Simplified circuit schematic of a single module of the fMUX TES readout. The colours indicate the components installed at different temperatures inside and outside the cryostat. The FPGA-based ICEboard sits at room temperature with the carrier and nuller DACs, the demod ADC, and the SQUID controller circuitry. The still stage at $\sim$700\,mK hosts the LC resonator board and the SQUID, while the TESs are located at the mixing chamber stage at around 12\,mK. The TESs are voltage biased with carrier tones and their respective currents summed up at the SQUID summing junction. When the digital active nulling (DAN) is active, these currents flow predominantly through the nuller line instead of the SQUID input coil, keeping the SQUID in the linear operating regime. Figure from\,\cite{Adamic:noise}.}
\label{fig:circuit}
\end{figure}

The fMUX readout system presented here is in principle very similar to the one developed for CMB experiments like POLARBEAR-2, Simons Array, SPT-3G and LiteBIRD (see Ref. \cite{Bender:2014nnc}), but adapted to cryogenic calorimetric experiments -- figure \ref{fig:circuit} shows the system configuration. It implements 10 detector channels, where each TES is placed in series with an LC circuit with a unique resonant frequency between 1-5 MHz. The readout electronics generate a comb of carrier tones, providing a voltage bias for each individual TES detector, which are kept at their operating points via electro-thermal feedback\,\cite{Irwin}. The resulting bias currents are summed together and inductively coupled to a DC-SQUID, which acts as a transimpedance amplifier. When energy is deposited in a TES, its resistance changes, creating an amplitude modulation of the current at its associated LC filter frequency, which is digitized and demodulated by the readout electronics. Our implementation of the fMUX TES readout, like its CMB predecessor, also makes use of Digital Active Nulling -- DAN \cite{deHaan:2012fx}, which dynamically injects a current along the nuller line that cancels the current flowing through the SQUID input coil, keeping the SQUID in the linear working regime. When DAN is active, all the science signal is therefore encoded in the nuller line (see figure \ref{fig:circuit}), while the demod line coming from the SQUID amplifier only contains the residuals (i.e., control error) that serve as an input to the DAN controller. The tone generation, demodulation, and the DAN loop are implemented digitally on McGill's FPGA-based "ICEboard" digital signal processing platform, described in more detail in section \ref{sec:warm}.

There has been some effort in the past to convert the CMB fMUX readout to a high-bandwidth version for superconducting gamma-ray spectrometers \cite{dreyer_fmux}. However, the hardware (both cold and warm) back then was much less sophisticated, the TES bias frequencies were lower, and most importantly, SQUID nulling was static, i.e. only the carrier tones were nulled out, while the sidebands were still carried over onto the demod line. Since the linearity of the detector readout is crucial for achieving the high energy resolution needed for CUPID, the DAN scheme in this modern implementation of high-bandwidth fMUX is a substantial step forward.

We tested the system in an Oxford Triton 400 dilution cryostat at UC Berkeley. The TESs employed for this demonstration are IrPt bilayer devices with a critical temperature, $T_c$, of 35\,mK or 55\,mK, described in\,\cite{Singh:2022rck}. They are deposited on high-purity Si wafers of 500~$\mu$m thickness and diced to $45\times45$~mm$^2$ in size. The TESs were installed on the mixing chamber stage at 12\,mK, while the LC filters and the SQUID are on two chips glued on a printed circuit board (PCB) on the still stage at slightly below 700\,mK -- see figure \ref{fig:circuit}. This cryogenic resonator board, together with the superconductive cabling, is described in more detail in section \ref{sec:cold} below.

\section{Cold components}
\label{sec:cold}

\subsection{Resonator board}
\begin{figure}[t]
    \centering
    \includegraphics[width=1.0\textwidth]{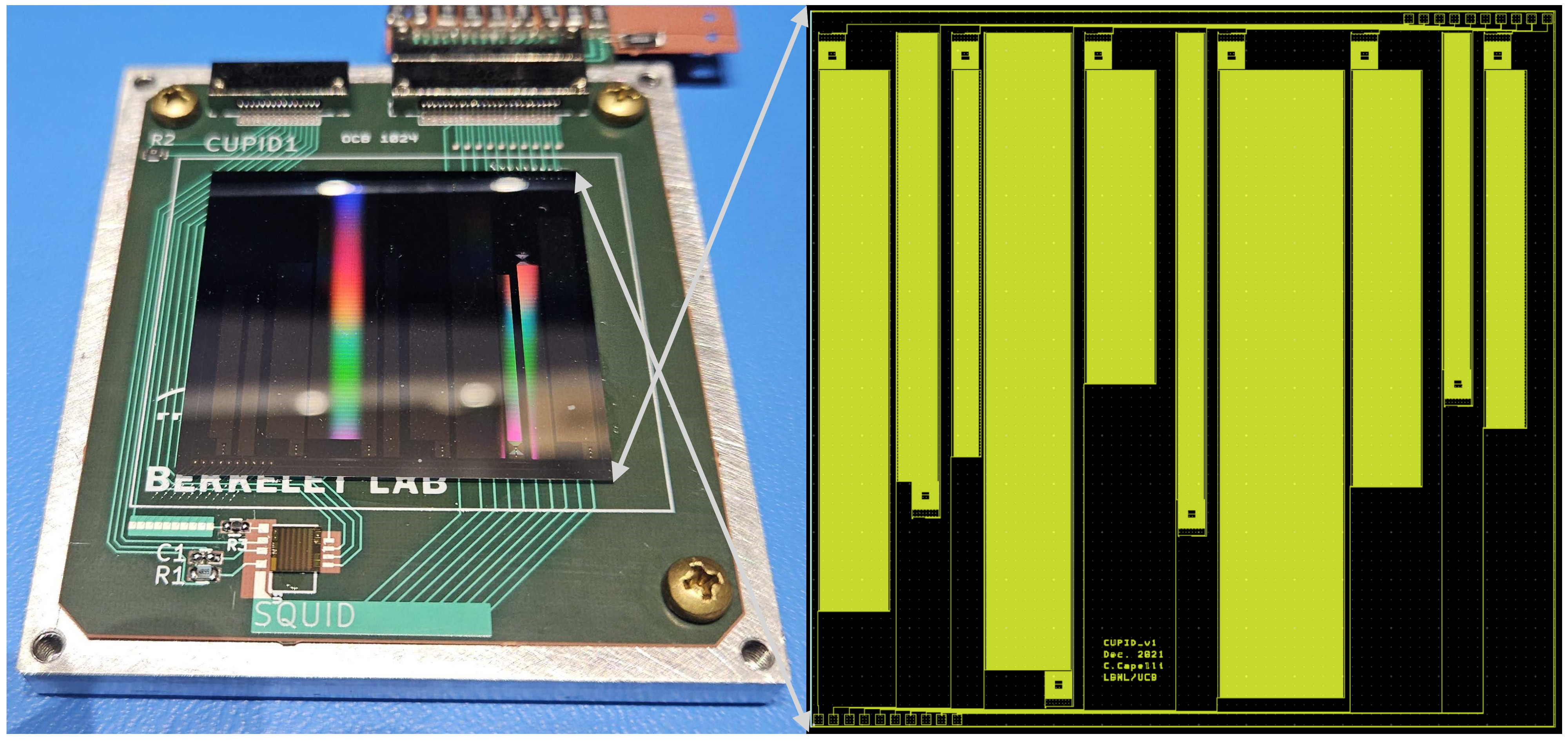}
    \caption{Left: cryogenic resonator board with the LC resonator chip in the center and the NIST SA13 SQUID at the bottom, together with the snubber (R1) and bias (R3) chip resistors. The board is placed in an aluminum box at the still stage at $\sim$700\,mK inside the cryostat. Right: zoom-in on the design of the LC resonator chip. The small squares are spiral inductors with constant value of $4$\,$\mu$H; we vary the size of the interdigitated (IDC) capacitors to get the desired resonance frequencies.}
    \label{fig:muxboard}
\end{figure}

The multiplexing PCB, also referred to as the resonator board (highlighted in yellow in figure~\ref{fig:circuit}), hosts both the resonator chip and a NIST SA13a DC-SQUID array \cite{sa13}. The two chips are attached to the PCB with rubber cement and connected to the board via 25\,$\mu$m aluminum wire bonds (see figure \ref{fig:muxboard}). The PCB is patterned with superconducting aluminum traces with copper added selectively to the pads, allowing for soldering to the readout wiring and other components with very low loss. A 5\,$\Omega$ snubber resistor (R1) is soldered in parallel with the SQUID input coil to remove non-linearities in the $V(\Phi)$ curves, which can occur when operating SQUID arrays at low temperatures\,\cite{HUBER_SQUID, Elleflot:2021hvi}. A 100\,$\Omega$ resistor (R2) serves as a heater in case it is necessary to warm the SQUID from its superconducting to its normal state. This can be useful if the SQUID has any remaining trapped flux acquired during cooldown. Lastly, a 20 m$\Omega$ bias resistor (R3) provides a frequency-independent voltage bias for the TESs.

The resonator board is enclosed in a high purity (5N) aluminum box that, in its superconducting state, protects the board from external magnetic fields. We wrapped the box in a layer of Metglas\textsuperscript{\textregistered} magnetic foil, providing further shielding from the external magnetic field. The assembly is thermalized to the still stage ($\sim$ 700 mK) through a copper rod.

Thermalizing TESs to the mixing chamber, while the resonator chip and the SQUID are mounted on the still stage, poses a challenge for cabling. Ideally these components would be as close to each other as possible to minimize attenuation and parasitic effects before the amplification of the signal. This ideal case is better realized in systems implemented for CMB telescopes, where both the TES and the readout SQUID can operate at the same temperature (about a couple hundreds of mK) \cite{Elleflot:2021hvi}. However, because this readout system is intended for low-background cryogenic calorimeters, the non-active material close to the detectors must be minimized and the detectors themselves must be as cold as possible (12 mK), where SQUIDs cannot reliably operate. To address this distance between the TESs and the SQUID, we use superconducting NbTi twisted pair cables to connect the TESs to the resonator board, where low-loss, fine-pitch Omnetics NanoD connectors make the connection to selective copper pads. On the TES side of the circuit, the connections are limited to NanoD connectors, Al solder pads, and Al wirebonds to keep the parasitic resistance as low as possible. The same choice of connectors and cables is also used to interface  the SQUID with room temperature control and readout. As seen later in section \ref{sec:measurements}, the stray inductance of this cabling will affect the performance of the system.

\subsection{Resonator chip design}
Our LC resonator chip is adapted from the POLARBEAR2 design \cite{Rotermund:2016uaz}, modified by the unique physics constraints of CUPID. The TESs developed for a CUPID-like experiment have an expected normal resistance of $\sim$1\,$\Omega$ and a time constant $\tau_{TES} \sim$ 50$\,\mu$s, roughly 200 times faster than POLARBEAR2. To ensure the stability of the calorimeter, $\tau_{TES}$ needs to exceed the electrical time constant of the RLC circuit, $\tau_{e}$, according to the relation\,\cite{Rotermund:2016uaz}:
\begin{equation}
    \tau_{TES} > 5.8 \cdot \tau_{e} \simeq 5.8 \cdot \frac{2L}{R_{TES}} \,.
\end{equation}
From this we obtain the desired value for the inductance of the resonator, $L <$ 4.3\,$\mu$H. For our design, we chose to fix the inductance value for all the resonators at $L=4$\,$\mu$H, and we vary the value of the coupled capacitance. This also keeps the bandwidth of each resonance constant. Considering a readout frequency of 1-5 MHz, we need capacitors on the order of $\mathcal{O}$(1\,nF).

The spacing between resonances follows the constraint of $\Delta f >$ 0.3\,MHz, in order to reduce the crosstalk, $XT$, to a sub-percentage level. The dominant component for the crosstalk is the leakage current between neighboring resonances, with a contribution given by\,\cite{Rotermund:2016uaz}
\begin{equation}
    XT_{lc} = \left( \frac{R_{TES}}{4\pi \Delta f L} \right)^2 \,,
\end{equation}
which is designed to be less than 0.4\,\% in our system. Another consideration when choosing the spacing of the resonances is minimizing the effect of intermodulation distortion (IMD) products between the carrier tones. Given that the distance to the second closest resonance is at least 0.6\,MHz with a $XT_{lc}<$0.1\,\%, we consider only the third order intermodulation effects between adjacent resonances. Thus, we space the resonant frequencies non-linearly to avoid the intermodulation distortion peaks, and with a resonance bandwidth of 40\,kHz. The selected resonant frequencies with the associated capacitances are reported in table \ref{tab:resonances}.

\begin{table}[thbp]
\centering
\caption{Resonances chosen for the design of the 10-channel resonator chip, with the corresponding capacitance needed when using an inductance of 4\,$\mu$H.\label{tab:resonances}}
\smallskip
\begin{tabular}{l|c|c|c}
\hline
No.&Resonance frequency [MHz]&Capacitance [nF] & Order on chip\\
\hline
1 & 1.472 & 2.932 & 3\\
2 & 1.772 & 2.017 & 6\\
3 & 2.144 & 1.378 & 9\\
4 & 2.444 & 1.060 & 2\\
5 & 2.816 & 0.799 & 5\\
6 & 3.166 & 0.652 & 8\\
7 & 3.488 & 0.521 & 1\\
8 & 3.788 & 0.441 & 4\\
9 & 4.160 & 0.366 & 10\\
10 & 4.460 & 0.318 & 7\\
\hline
\end{tabular}
\end{table}

The resonator chip is fabricated on a silicon substrate with a lithographed aluminum layer of planar, spiral inductors in series with interdigitated (IDC) capacitors. The resonators are placed next to each other in one row. To further reduce crosstalk, resonances close in frequency are spatially distant on the chip. The spatial ordering of the resonances on the chip is reported in the last column of table \ref{tab:resonances}, where the seventh resonance (3.488 MHz) corresponds to the leftmost capacitor in figure \ref{fig:muxboard}. In addition, adjacent resonators have their respective inductors on alternating sides of the capacitors to avoid mutual inductance.

The inductor is identical for each resonator and is made of 60 square turns with a trace gap and width of 4\,$\mu$m, resulting in a side dimension of 1400\,$\mu$m. The capacitors are made of two vertical pads with horizontal fingers in between, with a trace gap and width of 2\,$\mu$m. This is done to reduce the size of the capacitors so that the chip is as compact as possible. Each capacitor is connected to its inductor on one side and the bonding pad on the other, that will be connected to the TES. The size of the resonator chip is 38.4$\times$36.9\,mm$^2$, with an additional millimeter on each side for the dicing lines used during production. The dimensions of the entire resonator board are 70x60\,mm$^2$, which is small enough to fit on the still stage.

\section{Room temperature readout}
\label{sec:warm}

\subsection{The McGill ICEboard electronics}
\begin{figure}[t]
    \centering
    \includegraphics[width=0.7\textwidth]{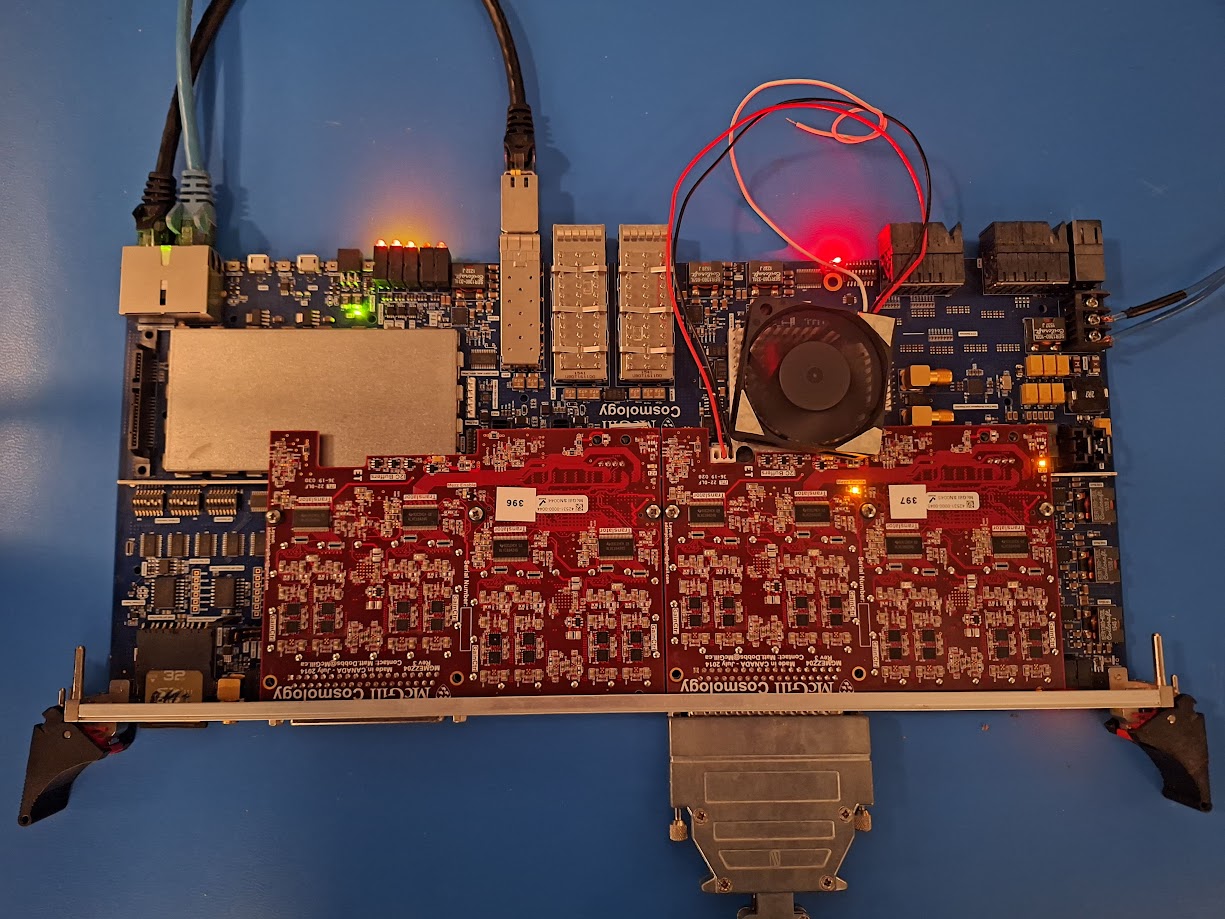}
    \caption{McGill's ICEboard warm readout electronics. The blue motherboard hosts the FPGA for digital processing, while the red mezzanine cards contain the ADCs and DACs. The DB-37 cable at the bottom connects to the SQUID Controller Board (not shown here), while the Ethernet cable at the top interfaces with the host PC. The board supports the readout of up to 8 fMUX modules, although we only use one in this work.}
    \label{fig:iceboard}
\end{figure}

For the digital frequency multiplexing control and readout of the cryogenic detectors, we are using the McGill Cosmology Instrumentation Laboratory's ICEboard electronics \cite{Bandura:2016dpm}, shown in figure~\ref{fig:iceboard}. The motherboard, originally developed for CMB experiments, is based on a high-performance Xilinx Kintex-7 FPGA, which runs the digital signal processing firmware for TES readout \cite{Bender:2014nnc}. This includes synthesizing and demodulating readout tones, the DAN control loop, and the downsampling and streaming of science data, which are sent out through a 1 Gb/s Ethernet link. The board also contains an ARM processor running a Linux operating system, which serves as a low-level interface between the FPGA and the host computer.

The digital waveforms are converted to analog signals and the analog signals returning from the cryostat are digitized using 16-bit 20 Msps ADCs and DACs on the two custom FPGA mezzanine cards (FMCs) mounted on the ICEboard. The mezzanines connect to a separate SQUID Controller board \cite{Bender:2014nnc}, also made by McGill, connecting to the cryostat. The hardware supports the readout of up to 8 modules (each containing a SQUID and an array of detectors), although only one is currently enabled by the CUPID firmware due to bandwidth limitations specific to the present configuration of the ARM's Ethernet transceiver, discussed in more detail in section \ref{subsec:firmware}. The Iceboard has high-speed serial links available (SFP or QSFP directly from the FPGA), so this will not be an issue in the long term.

\subsection{New high-bandwidth fMUX firmware}
\label{subsec:firmware}
30 ICE boards are currently deployed at the South Pole Telescope (SPT) in Antarctica, where they read out approximately 16,000 TES detectors for the SPT-3G experiment \cite{SPT-3G:2014dbx, SPT-3G:2021vps}, scanning the CMB sky. For CUPID TES readout, we are using the exact same warm hardware as for SPT-3G; however, the science requirements are very different. While the SPT slowly scans the sky, requiring a low sampling rate, CUPID must read out its detectors much faster to avoid $2\nu\beta\beta$ pileup events. Therefore, new fast readout firmware was designed at McGill to meet its science goals, specifically for CUPID. We present its inner workings below.

\begin{figure}[t]
    \centering
    \includegraphics[width=1.0\textwidth]{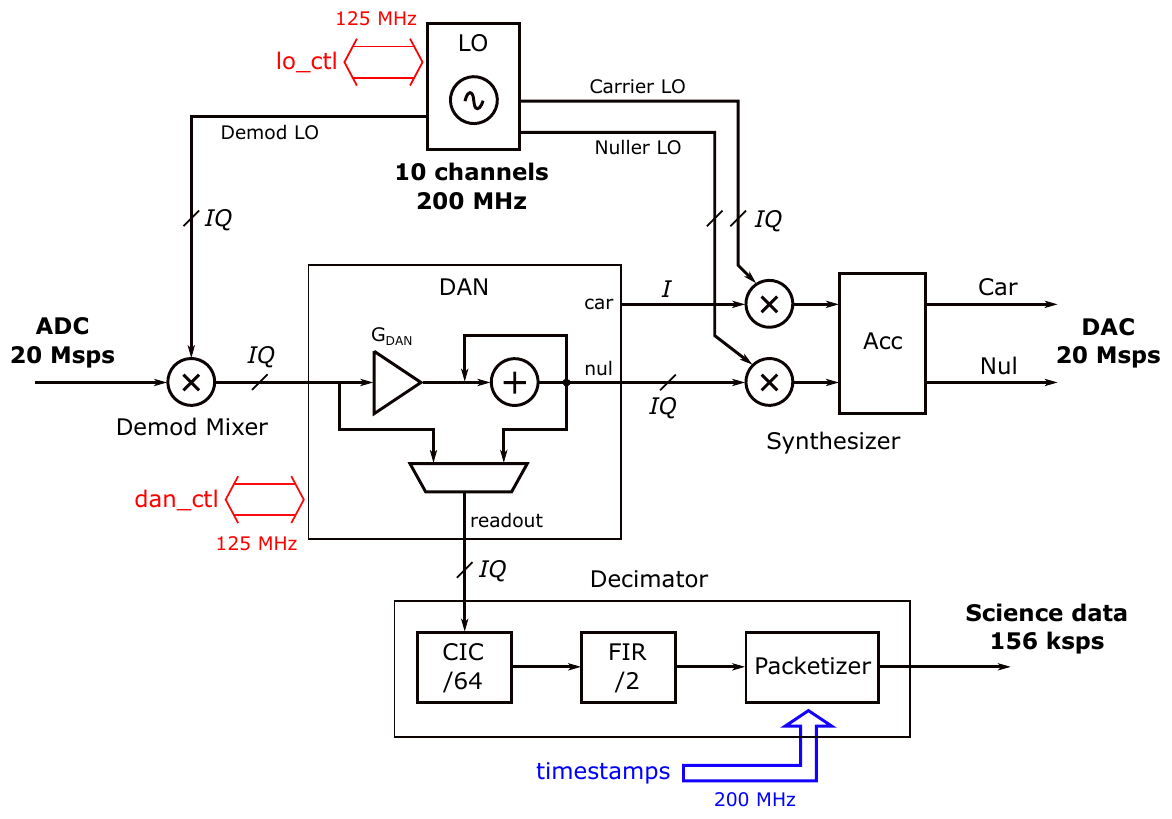}
    \caption{Core architecture of the high-bandwidth, 10-channel fMUX firmware module for CUPID, clocked at 200 MHz. The main difference compared to McGill's SPT firmware \cite{Bender:2014nnc} is the absence of polyphase filter banks (PFBs) and a new decimation path. The DAN loop remains the same, albeit it runs at a higher sampling rate. The operation of the firmware is controlled via 125 MHz control interfaces, marked in red, which ultimately connect to the ARM processor through an on-board SPI link. There is also a Global Positioning System (GPS)-derived input in IRIG-B format, which attaches timestamps to the data packets (in blue). Timestamps have a programmable delay (not shown here), which aligns them with the correct sampled data.}
    \label{fig:firmware}
\end{figure}

While the SPT firmware \cite{Bender:2014nnc} is optimized for high multiplexing ratios at low sample rates, CUPID does the opposite. The new firmware increases the output data rate to 156 ksps, up by three orders of magnitude compared to 153 sps for SPT-3G. On the other hand, the fMUX ratio is reduced to 10x, down from 128x for SPT-3G. The SPT firmware achieves such high multiplexing factors through the use of polyphase filter banks (PFBs), a very efficient way of splitting the band into multiple narrow channels. Since the requirements for CUPID are the opposite (only a few very wide channels), the PFBs have been removed, significantly reducing the firmware footprint of the design. The architecture of the new firmware is shown in figure \ref{fig:firmware}.

The firmware signal processing works as follows. Raw data comes in from the analog-to-digital converter (ADC) at 20 Msps and is quadrature demodulated to baseband by mixing it with the local oscillator (LO) signal, which generates the tones for all 10 channels using Direct Digital Synthesis (DDS). Channel processing is time-multiplexed, corresponding to a system clock of 200 MHz (200 MHz/10 = 20 Msps). Turning to the synthesis side on the right of figure \ref{fig:firmware}, a series of carrier and nuller waveforms are generated by mixing their baseband amplitudes with the LO signals. The accumulator sums all of them together, creating a frequency comb at 20 Msps, which feeds the carrier and nuller DACs (digital-to-analog converters).

The decimator downsamples the 20 Msps data to 312.5 ksps (/64) with a 6-stage cascaded integrator-comb (CIC) filter \cite{Hogenauer_1981} with an anti-aliasing rejection of more than -60 dB. Since CICs are known to have a significant magnitude droop in the passband, a compensating FIR (finite impulse response) decimator (/2) follows, flattening the passband for the final science output data rate of 156.25 ksps. The channelized data are packetized and sent off the FPGA to the ARM processor over Ethernet.

At the core of the readout module is the DAN control loop \cite{deHaan:2012fx}, which actively adjusts the nuller amplitude to cancel any signal at the input (see also figure \ref{fig:circuit}). When DAN is enabled, the nuller line encodes the science data, so it is the one being sampled and packetized instead of the demod line. The DAN module is an integral controller with an adjustable digital gain $G_{DAN}$, which sets the aggressiveness of the feedback. The total feedback loopgain, $K_0$, is the product of the set gain $G_{DAN}$ and all other gains in the external nuller-demod circuit loop, here denoted as "loopgain" $LG$.
\begin{equation}
    K_0 = G_{DAN} \times LG
\end{equation}
The higher the feedback gain $K_0$, the faster the nuller will respond to changes at the SQUID input coil. Specifically, the signal bandwidth is linearly proportional to $K_0$, and in general, we would like the bandwidth to be as high as possible. However, there is an upper limit to $K_0$, above which \textit{antinulling} occurs and the control loop becomes unstable, a phenomenon discussed in detail in Ref. \cite{SmecherDAN}. Two factors are limiting this. The first is the loop latency, which was estimated to be $\sim 2.2 \, \mu$s for this system, a significant improvement from 14 $\mu$s for the SPT firmware (due to removing the polyphase filter banks -- PFBs). The second limiting factor is due to loopgain nonuniformity, which was observed during operation and is unique to high-bandwidth TES readout. Since the readout Nyquist frequency is high (78 kHz), the resonator impedance varies significantly within it, and so does the loopgain $LG$ (due to the varying nuller current sharing factor -- see \cite{Adamic:noise}). In other words, $LG$ increases at frequencies away from the center of the resonance. This causes the points at the edge of the Nyquist region to enter the antinulling regime. Since only a single value of $G_{DAN}$ can be programmed per channel, the entire channel's digital gain must be set more conservatively. Using the current firmware and readout configuration, the highest attained safe DAN readout bandwidth is around 3 kHz. For a first-order low pass filter, the correlation between fastest possible risetime and its bandwidth is $t_r \approx 0.35/BW$, so a 3 kHz cutoff corresponds to detector risetimes of $\sim 120 \, \mu$s. If there was a need to improve this, one could either widen the LC resonances (which will adversely affect crosstalk), implement a higher-order DAN controller, or reduce the output sample rate of the system. In addition, digital latency could be reduced by optimizing the DAN firmware, which still includes a lot of unnecessary components from SPT-3G; for example, SRON's BBFB \cite{denHartogBBFeedbackSRON} has sub-$\mu$s latency and can achieve an order of magnitude higher feedback loop bandwidth.

The single readout module CUPID firmware was implemented on the Iceboard's Kintex-7 420T FPGA using Xilinx Vivado 2021.2 ML. The reported firmware utilization is 4.0 \% of Lookup Tables (LUTs), 3.1 \% of Registers (total Slice utilization 7.2 \%), 3.7 \% of Block Random-Access Memory (BRAMs) and 2.7 \% of Digital Signal Processing (DSP) blocks with an estimated on-chip power of merely 2 W. Therefore, the design is easily scalable up to 8 modules, which is the physical number of hardware readout modules on the Iceboard.

The issue for the time being is the ARM-based network bandwidth, which cannot handle that many UDP packets. A single IQ datapoint is a 64-bit word, times 10 channels at 156250 sps equals to 100 Mbit/s, not taking into account the structure of our UDP packets which adds another 10\% overhead, so a total datarate of $\sim$110 Mbit/s per module. The packets are sent from the FPGA to the on-board ARM processor through a short UTP loopback, where a special process runs that routes the packets to the host computer via a separate 1.0 Gbps Ethernet port. The reason behind this configuration is that the ARM Linux system hosts a set of C-based algorithms which serve as a low-level control interface between the firmware and the host computer, and sometimes these algorithms need access to output data. This process-based packet routing has a bottleneck at around 150 Mbps, limiting the firmware to only one active 10x module. The solution to this is known and successfully implemented in McGill's RF-ICE \cite{Rouble:RF-ICE} GHz multiplexing Iceboard system, where the data rates can go above 500 Mbps, which would enable the implementation of 4 modules. However, for the full 8-module system, the 1.0 Gbps Ethernet interface hard limit will become a problem, prompting the transition to on-board 10 Gbps links \cite{Bandura:2016dpm}, used by default on Iceboards deployed at the CHIME radio telescope \cite{CHIME:2022dwe}.

We also successfully implemented and tested a 15x frequency multiplexing version of the firmware, using various modifications that enabled increasing the clock frequency to 300 MHz. However, similarly to the 8-module readout, it is limited by network bandwidth constraints. In the future, it is essential to rethink how we offload data from CUPID Iceboards to unlock the full scalability of this high-bandwidth TES fMUX readout firmware. Using the high-speed interface and a new software control system would enable scaling to 8 modules with 15x multiplexing, or 120 detectors per board.

\section{Multiplexing demonstration}
\label{sec:measurements}

To validate and characterize the system, we first carried out measurements replacing the TES array with a board with a "dummy payload" of 10 surface mount (SMD) 0.5\,$\Omega$ resistors, attached to the resonator board and thermalized to the still stage at 700 mK. This allowed us to perform the basic multiplexing and noise characterization \cite{Adamic:noise} of the system first without the complications related to operating TES detectors.

\begin{figure}[t]
    \centering
    \includegraphics[width=\textwidth]{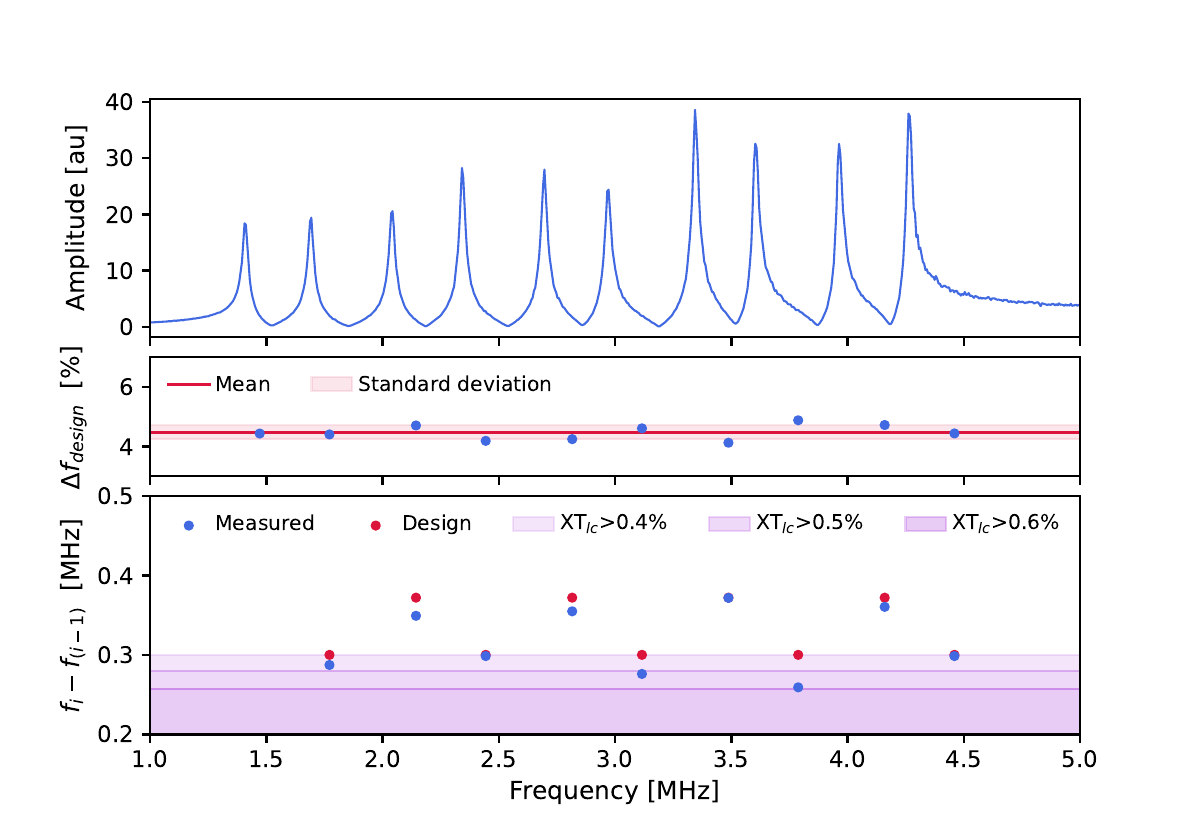}
    \caption{Top panel: DAN network analysis of the system with 0.5\,$\Omega$ SMD resistors. All 10 resonances are visible, with peaks corresponding to chosen detector bias points. Middle panel: shift in frequency between the measured resonance values (in the top panel) and the design ones, with the point \textit{x}-axis corresponding to design resonance frequencies. Mean and standard deviation are shown by the solid line and the shaded red region, respectively. Bottom panel: the distance between neighboring resonances for measured (blue) and design (red) values -- note that the red and blue dots overlap in three cases. The contribution of the leakage cross-talk is overlaid in shades of purple.}
    \label{fig:resonances}
\end{figure}

The top panel of figure \ref{fig:resonances} shows the network analysis of the setup from 1 to 5 MHz, performed with the Iceboard. The plot is showing the "DAN network analysis", which is produced by physically measuring the carrier transfer function and dividing it by the nuller transfer function. This eliminates the impedance of the SQUID input coil from the expression, corresponding to the true comb admittance when DAN is operating.\,\cite{Joshua_PhD} The TESs (or in this case regular resistors) are subsequently biased at the maxima of such network analysis. All ten resonances are visible, with an average shift towards lower frequencies of (4.5$\pm$0.2)\,\% from the design (middle panel). This corresponds to an increase of the LC constant of the resonators of (9.6$\pm$0.5)\,\%, indicating the presence of parasitic components, mostly inductance from the TES-SQUID NbTi wiring. The shift in frequency changes the relative distance among the resonances, potentially affecting crosstalk, $XT_{lc}$. The bottom panel shows the design and measured resonance frequencies and different $XT_{lc}$ levels. While most resonance pairs still satisfy the design requirement of $XT_{lc} \le 0.4$\,\%, three pairs have a $XT_{lc} \le 0.6$\,\%.

\begin{figure}[t]
    \centering
    \includegraphics[width=0.8\textwidth]{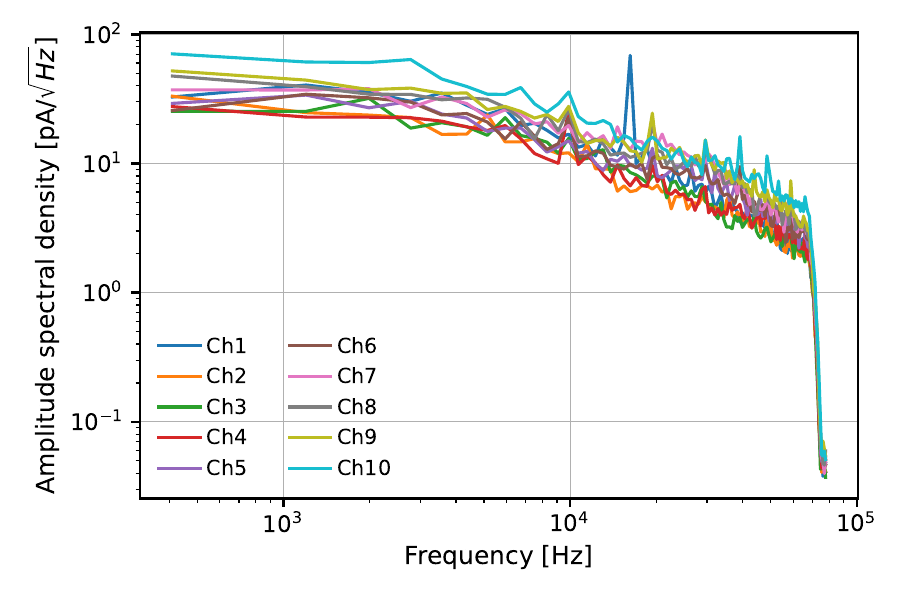}
    \caption{Noise spectra of the 10 active channels with DAN enabled, measured as current on the nuller line -- see figure \ref{fig:circuit}. The 3 kHz white noise roll-off is clearly visible, demonstrating the required readout bandwidth. The spikes in the high frequency part of the spectrum are IMD products between the readout tones and the 20 MHz ADC/DAC clock. For the discussion of white noise amplitudes, see Ref. \cite{Adamic:noise}.}
    \label{fig:noise_spectra}
\end{figure}

We have also measured the readout noise of the system with DAN enabled, corresponding to TES readout currents on the nuller line -- see figure \ref{fig:noise_spectra}. The spectra have a noise roll-off at 3 kHz, matching the bandwidth of the DAN loop discussed earlier. This demonstrates the readout bandwidth corresponding to a $120 \, \mu$s signal risetime detection capability, meeting the design requirement of $\sim$170 $\mu$s. The measured white noise levels across the DAN bandwidth were 20 -- 50 $\frac{pA}{\sqrt{Hz}}$ at the SQUID summing junction, depending on the bias frequency (noise generally rises with frequency in this system); the readout noise, excluding the Johnson noise contribution from TES resistors, was estimated to 15 -- 40 $\frac{pA}{\sqrt{Hz}}$. We kindly refer the reader to \cite{Adamic:noise}, where we provide a detailed investigation of the readout noise. To summarize, the contributions of typical noise sources in the system, like the SQUID, the first stage amplifier and various resistors, are significantly amplified in the CUPID case owing to two main effects. The first one is nuller current sharing, which happens because of low TES operating resistance compared to the SQUID input impedance \cite{Adamic:noise}. The second is SQUID output filtering, due to long wiring in the CUPID cryostat having large capacitance, which forms a low-pass filter at the output of the SQUID and attenuating the signal, therefore boosting warm readout noise with DAN feedback enabled \cite{Adamic:noise}. Both of these noise-amplifying effects are well-understood in the CMB fMUX community and have been successfully addressed in CMB systems \cite{Elleflot:2021hvi, Joshua_PhD}. For CUPID, the solution towards low-noise readout is to either increase the TES operating resistance or use a lower input inductance SQUID, in order to decrease current sharing, and to employ shorter or low-capacitance wiring at the output of the SQUID, to minimize signal attenuation \cite{Adamic:noise}.

In figure \ref{fig:noise_spectra}, we notice some intermodulation distortion (IMD) products in the spectra, which form between the readout tones and the 20 MHz sampling clock, not considered at the design stage. These spikes are not problematic for a CUPID-type experiment, since the pulse information is encoded in the entire readout bandwidth. With additional tone frequency scheduling in future design iterations, most IMD spikes can be avoided, and any remaining ones can be notched out in post-processing of the data.

\begin{figure}[t]
   \begin{minipage}[t]{0.46\textwidth}
     \includegraphics[height=\textwidth, valign=t]{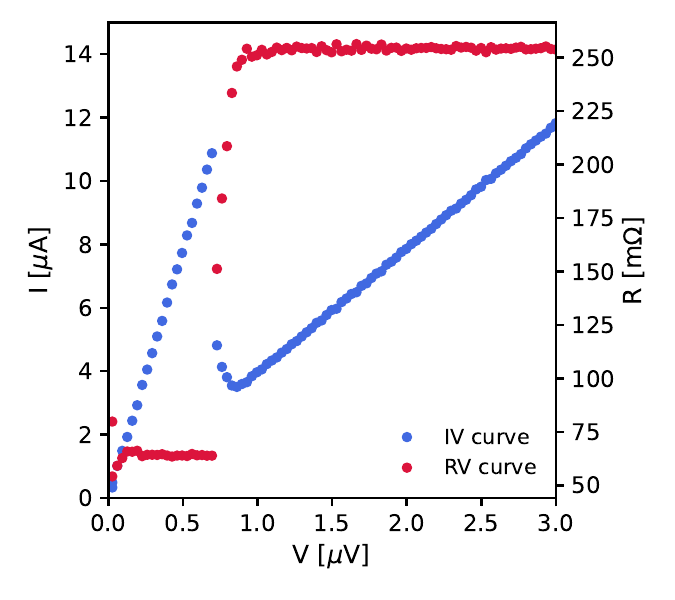}
     \caption{Current (I) and measured resistance (R) of the fifth TES as a function of bias voltage (V) as the bolometers are lowered into transition. Residual parasitic resistance (about 70~m$\Omega$) is clearly visible in the superconducting state.}
     \label{fig:loadcurves}
   \end{minipage}
   \hfill
   \begin{minipage}[t]{0.46\textwidth}
     \includegraphics[width=\textwidth, valign=t]{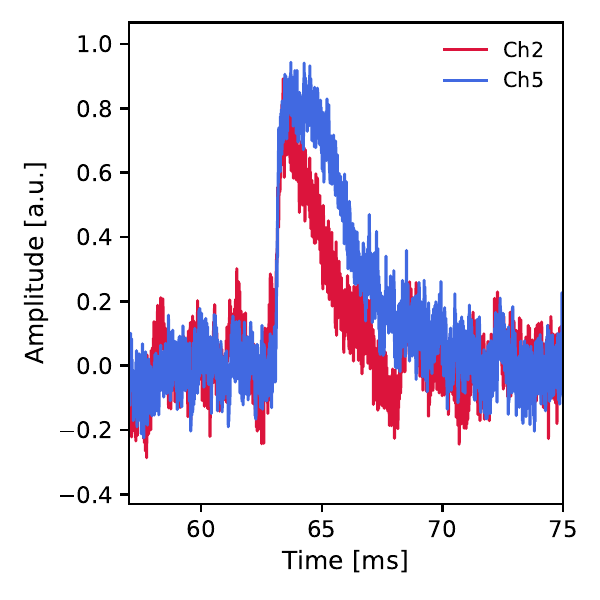}
     \caption{Example of a short LED pulse picked up by two TES devices via a semi-transparent optic fiber inside the cryostat.}
     \label{fig:ledpulses}
   \end{minipage}
\end{figure}

% TES section
Lastly, we tested the system with an array of 9 TESs installed at 12 mK, developed specifically for CUPID in \cite{Singh:2022rck} and briefly introduced in section \ref{sec:fmux}. We performed a DAN network analysis at 20\,mK and 75\,mK (below and above transition) to visualize the admittance variation of the resonances between the superconductive and normal regimes. We observed 6 devices transitioning. The DAN was enabled for all TES channels above their transition temperatures, and high bias voltages were programmed to keep them in the normal regime. After the mixing stage was lowered below $T_c$, the bias voltages were slowly reduced to lower the TESs into transition -- see figure \ref{fig:loadcurves}. We could estimate a series parasitic resistance between 50\,m$\Omega$ and 75\,m$\Omega$ on the TES lines from these load curves, most likely originating from connector contact impedance, dielectric losses in silicon and inductor losses through eddy currents from inductors' magnetic fields. This resistance affects the shape of the pulses; in particular, it affects the thermal response of the detector. An increase in the parasitic resistance decreases the loopgain of the electro-thermal feedback, which is inversely proportional to the effective time response of the TES, thus increasing the risetime and reducing the pileup discrimination capabilities of the detector. This effect was studied for CMB and POLARBEAR experiments and it is detailed in \cite{Elleflot:2019lbn}.

To demonstrate calorimetric reconstruction of optical photons with the developed fMUX readout, as is necessary to discriminate backgrounds in CUPID, we routed a cryogenic diffusive optical fiber down to the TES array at 12 mK. The fiber mates to the vacuum feedthrough at 300\,K, which connects to another segment of fiber leading to an LED at room temperature. A signal generator drives the LED with a square pulse train, generating optical photons with an intensity proportional to the pulse width. The pulse width of the signal generator was tuned to keep the TESs below their saturation limit. We acquired data on two TES devices, shown in figure \ref{fig:ledpulses}, where we can see demodulated time streams showing coincident pulses picked up by both TESs.

\section{Conclusion and outlook}
\label{sec:conclusion}

We developed a frequency domain multiplexing readout system for TESs with a signal bandwidth in the kHz range, as required by the next generation of cryogenic calorimetric double beta decay experiments. The readout system is adapted from CMB experiments, namely POLARBEAR2 and SPT-3G. The higher readout bandwidth required a new design of the LC resonator board and a new firmware for the FPGA signal processing motherboard. The installation in the cryostat and the cabling were modified to reduce the heat load on the mixing chamber at $\sim$12\,mK and to minimize the material close to the detector, as required by rare event search experiments. The new system was successful in reading out multiplexed devices, which respond to light pulses. Residual parasitic impedance was measured, originating from cables and connectors. Although some further optimization is required, the system meets the physics requirements for next-generation rare-event cryogenic calorimeters, is scalable to higher channel counts, and matches the state-of-the-art fMUX microcalorimeter readout systems in terms of multiplexing ratio, bandwidth and readout noise. The multiplexed readout developed herein facilitates the deployment of thousands of TES sensor channels, a scale crucial for future tonne-scale detectors.
%\appendix
%\section{Some title}

%We suggest not to abbreviate: ``section'', ``appendix'', ``figure''
%and ``table'', but ``eq.'' and ``ref.'' are welcome. Also, please do
%not use \texttt{\textbackslash emph} or \texttt{\textbackslash it} for
%latin abbreviaitons: i.e., et al., e.g., vs., etc.

\acknowledgments
This work was supported by the U.S. Department of Energy (DOE) Office of Science, Offices of Nuclear Physics and High Energy Physics, under Contract Nos. DE-AC02-05CH11231 and DE-AC02-06CH11357, and the DOE Office of Science, Office of Nuclear Physics under Contract Nos. DE-FG02-00ER41138 and DE-SC0020423. J. Camilleri is supported by the DOE Office of Science Graduate Student Research (SCGSR) program, administered by the Oak Ridge Institute for Science and Education for the DOE under contract number DE-SC0014664.
The McGill team acknowledges funding from the Natural Sciences and Engineering Research Council of Canada (NSERC), the Canadian Institute for Advanced Research (CIFAR), and the Canada Research Chairs (CRC) program.

%\paragraph{Note added.} This is also a good position for notes added
%after the paper has been written.

% Bibliography

%% [A] Recommended: using JHEP.bst file
\bibliographystyle{JHEP}
\bibliography{biblio.bib}

@article{Singh:2022rck,
    author = "Singh, V. and others",
    title = "{Large-area photon calorimeter with Ir-Pt bilayer transition-edge sensor for the CUPID experiment}",
    doi = "10.1103/PhysRevApplied.20.064017",
    journal = "Phys. Rev. Applied",
    volume = "20",
    number = "6",
    pages = "064017",
    year = "2023",}

@article{Dobbs_fmux,
author = {Dobbs,M. A.  and others},
title = {Frequency multiplexed superconducting quantum interference device readout of large bolometer arrays for cosmic microwave background measurements},
journal = {Rev. Sci. Instrum.},
volume = {83},
number = {7},
pages = {073113},
year = {2012},
doi = {10.1063/1.4737629},
URL = {https://doi.org/10.1063/1.4737629}
}

@article{POLARBEAR:2015jaw,
    author = "Hattori, K. and others",
    editor = "Camus, Philippe and Juillard, Alexandre and Monfardini, Alessandro",
    collaboration = "POLARBEAR",
    title = "{Development of readout electronics for POLARBEAR-2 Cosmic Microwave Background experiment}",
    doi = "10.1007/s10909-015-1448-x",
    journal = "J. Low Temp. Phys.",
    volume = "184",
    number = "1-2",
    pages = "512--518",
    year = "2016"
}

@incollection{Irwin,
author="Irwin, K.D.
and Hilton, G.C.",
editor="Enss, Christian",
title="Transition-Edge Sensors",
bookTitle="Cryogenic Particle Detection",
year="2005",
publisher="Springer Berlin Heidelberg",
address="Berlin, Heidelberg",
pages="63--150",
isbn="978-3-540-31478-3",
doi="10.1007/10933596_3",
url="https://doi.org/10.1007/10933596_3",
}

@article{Pyle:2006rp,
    author = "Pyle, M. and Brink, P. L. and Cabrera, B. and Castle, J. P. and Colling, P. and Chang, C. L. and Cooley, J. and Lipus, T. and Ogburn, R. W. and Young, B. A.",
    editor = "Ohkubo, M. and Mitsuda, K. and Takahashi, H.",
    title = "{Quasiparticle propagation in aluminum fins and tungsten TES dynamics in the CDMS ZIP detector}",
    doi = "10.1016/j.nima.2005.12.022",
    journal = "Nucl. Instrum. Meth. A",
    volume = "559",
    pages = "405--407",
    year = "2006"
}

@article{Rotermund:2016uaz,
    author = "Rotermund, K. and Barch, B. and Chapman, S. and Hattori, K. and Lee, A. and Palaio, N. and Shirley, I. and Suzuki, A. and Tran, C.",
    title = "{Planar Lithographed Superconducting LC Resonators for Frequency-Domain Multiplexed Readout Systems}",
    doi = "10.1007/s10909-016-1554-4",
    journal = "J. Low Temp. Phys.",
    volume = "184",
    number = "1-2",
    pages = "486--491",
    year = "2016"
}

@article{SPT-3G:2014dbx,
    author = "Benson, B. A. and others",
    collaboration = "SPT-3G",
    title = "{SPT-3G: A Next-Generation Cosmic Microwave Background Polarization Experiment on the South Pole Telescope}",
    reportNumber = "FERMILAB-CONF-14-197-AE",
    doi = "10.1117/12.2057305",
    journal = "Proc. SPIE Int. Soc. Opt. Eng.",
    volume = "9153",
    pages = "91531P",
    year = "2014"
}

@article{SPT-3G:2021vps,
    author = "Sobrin, J. A. and others",
    collaboration = "SPT-3G",
    title = "{The Design and Integrated Performance of SPT-3G}",
    reportNumber = "FERMILAB-PUB-21-291-AE",
    doi = "10.3847/1538-4365/ac374f",
    journal = "Astrophys. J. Supp.",
    volume = "258",
    number = "2",
    pages = "42",
    year = "2022"
}

@article{deHaan:2012fx,
    author = "de Haan, Tijmen and Smecher, Graeme and Dobbs, Matt",
    title = "{Improved Performance of TES Bolometers using Digital Feedback}",
    doi = "10.1117/12.925658",
    journal = "Proc. SPIE Int. Soc. Opt. Eng.",
    volume = "8452",
    pages = "84520E",
    year = "2012"
}

@article{SmecherDAN,
title={Digital Active Nulling for Frequency-Multiplexed Bolometer Readout: Performance and Latency}, 
author={Graeme Smecher and Tijmen de Haan and Matt Dobbs and Joshua Montgomery},
year={2022},
eprint={2207.11377},
archivePrefix={arXiv},
primaryClass={astro-ph.IM},
url={https://arxiv.org/abs/2207.11377}, 
}

@ARTICLE{Hogenauer_1981,
  author={Hogenauer, E.},
  journal={IEEE Trans. Acoust., Speech, Signal Process.}, 
  title={An economical class of digital filters for decimation and interpolation}, 
  year={1981},
  volume={29},
  number={2},
  pages={155-162},
  doi={10.1109/TASSP.1981.1163535}
}

@article{Bender:2014nnc,
    author = "Bender, Amy N. and others",
    title = "{Digital frequency domain multiplexing readout electronics for the next generation of millimeter telescopes}",
    doi = "10.1117/12.2054949",
    journal = "Proc. SPIE Int. Soc. Opt. Eng.",
    volume = "9153",
    pages = "91531A",
    year = "2014"
}

@article{Bandura:2016dpm,
    author = "Bandura, K. and others",
    title = "{ICE: a scalable, low-cost FPGA-based telescope signal processing and networking system}",
    doi = "10.1142/S2251171716410051",
    journal = "J. Astron. Inst.",
    volume = "05",
    number = "04",
    pages = "1641005",
    year = "2017"
}

@article{CUPID_1T,
    author = "Armatol, A. and others",
    collaboration = "CUPID",
    title = "{Toward CUPID-1T}",
    eprint = "2203.08386",
    archivePrefix = "arXiv",
    primaryClass = "nucl-ex",
    month = "3",
    year = "2022"
}

@article{CUPID:2025avs,
    author = "Alfonso, K. and others",
    collaboration = "CUPID",
    title = "{CUPID, the CUORE upgrade with particle identification}",
    doi = "10.1140/epjc/s10052-025-14352-1",
    journal = "Eur. Phys. J. C",
    volume = "85",
    number = "7",
    pages = "737",
    year = "2025",
    note = "[Erratum: Eur.Phys.J.C 85, 1346 (2025)]"
}

@article{Alduino:2019xia,
    author = "Alduino, C. and others",
    title = "{The CUORE cryostat: An infrastructure for rare event searches at millikelvin temperatures}",
    doi = "10.1016/j.cryogenics.2019.06.011",
    journal = "Cryogenics",
    volume = "102",
    pages = "9--21",
    year = "2019"
}

@article{Chernyak:2016aps,
    author = "Chernyak, D. M. and others",
    title = "{Rejection of randomly coinciding events in Li$_2^{100}\mathrm{MoO}_4$ scintillating bolometers using light detectors based on the Neganov\textendash{}Luke effect}",
    doi = "10.1140/epjc/s10052-016-4565-z",
    journal = "Eur. Phys. J. C",
    volume = "77",
    number = "1",
    pages = "3",
    year = "2017"
}

@article{Rouble:RF-ICE,
    author = "M. Rouble and G. Smecher and A. Anderson and P. S. Barry and K. Dibert and M. Dobbs and K. S. Karkare and J. Montgomery",
    title = "{RF-ICE: large-scale gigahertz readout of frequency-multiplexed microwave kinetic inductance detectors}",
    doi = "10.1117/12.2630286",
    journal = "Proc. SPIE Int. Soc. Opt. Phot.",
    volume = "12190",
    pages = "1219024",
    year = "2022",
    booktitle = {Millimeter, Submillimeter, and Far-Infrared Detectors and Instrumentation for Astronomy XI},
    editor = {Jonas Zmuidzinas and Jian-Rong Gao},
}

@article{CHIME:2022dwe,
    author = "Amiri, Mandana and others",
    collaboration = "CHIME",
    title = "{An Overview of CHIME, the Canadian Hydrogen Intensity Mapping Experiment}",
    doi = "10.3847/1538-4365/ac6fd9",
    journal = "Astrophys. J. Supp.",
    volume = "261",
    number = "2",
    pages = "29",
    year = "2022"
}

@article{sa13,
author={Silva-Feaver, Maximiliano
and Arnold, Kam
and Barron, Darcy
and Denison, Edward V.
and Dobbs, Matt
and Groh, John
and Hilton, Gene
and Hubmayr, Johannes
and Irwin, Kent
and Lee, Adrian
and Vale, Leila R.},
title="{Comparison of NIST SA13a and SA4b SQUID Array Amplifiers}",
journal={J. Low Temp. Phys.},
year={2018},
month={Nov},
day={01},
volume={193},
number={3},
pages={600-610},
issn={1573-7357},
doi={10.1007/s10909-018-2052-7},
url={https://doi.org/10.1007/s10909-018-2052-7}
}

@ARTICLE{HUBER_SQUID,

  author={Huber, M.E. and Neil, P.A. and Benson, R.G. and Burns, D.A. and Corey, A.M. and Flynn, C.S. and Kitaygorodskaya, Y. and Massihzadeh, O. and Martinis, J.M. and Hilton, G.C.},
  journal={IEEE Trans. Appl. Supercond.}, 
  title="{DC SQUID series array amplifiers with 120 MHz bandwidth}", 
  year={2001},
  volume={11},
  number={1},
  pages={1251-1256},
  doi={10.1109/77.919577}}

@article{Elleflot:2021hvi,
    author = "Elleflot, Tucker and others",
    title = "{Low Noise Frequency-Domain Multiplexing of TES Bolometers Using SQUIDs at Sub-Kelvin Temperature}",
    doi = "10.1007/s10909-022-02796-8",
    journal = "J. Low Temp. Phys.",
    volume = "209",
    number = "3-4",
    pages = "693--701",
    year = "2022"
}

@ARTICLE{Adamic:noise,
  author={Adamič, Michel and Camilleri, Joseph and Capelli, Chiara and Dobbs, Matt and Elleflot, Tucker and Kolomensky, Yury G. and Mayer, Daniel and Montgomery, Joshua and Novosad, Valentine and Singh, Vivek and Smecher, Graeme and Suzuki, Aritoki and Welliver, Bradford},
  journal={IEEE Trans. Nucl. Sci.}, 
  title="{Readout noise of digital frequency multiplexed TES detectors for CUPID}", 
  year={2025},
  volume={},
  number={},
  pages={1-1},
  doi={10.1109/TNS.2025.3638275}}

@phdthesis{Joshua_PhD,
  author = {Joshua Montgomery},
  title = {Digital Frequency Domain Multiplexing readout: design and performance of the SPT-3G instrument and LiteBIRD satellite readout},
  school = {McGill University},
  year = {2020},
  address = {Montreal, QC, Canada},
  month = {November},
}

@ARTICLE{dreyer_fmux,
  author={Dreyer, Jonathan G. and Arnold, Kam and Lanting, Trevor M. and Dobbs, Matt A. and Friedrich, Stephan and Lee, Adrian T. and Spieler, Helmuth G.},
  journal={IEEE Trans. Appl. Supercond.}, 
  title={Frequency-Domain Multiplexed Readout for Superconducting Gamma-Ray Detectors}, 
  year={2007},
  volume={17},
  number={2},
  pages={633-636},
  doi={10.1109/TASC.2007.898249}}

@article{Barabash:2020nck,
    author = "Barabash, Alexander",
    title = "{Precise Half-Life Values for Two-Neutrino Double-{\ensuremath{\beta}} Decay: 2020 Review}",
    doi = "10.3390/universe6100159",
    journal = "Universe",
    volume = "6",
    number = "10",
    pages = "159",
    year = "2020"
}

@article{BICEP:2023jup,
    author = "Fatigoni, S. and others",
    collaboration = "BICEP",
    title = "{Results and Limits of Time-Division Multiplexing for the BICEP Array High-Frequency Receivers}",
    doi = "10.1007/s10909-024-03100-6",
    journal = "J. Low Temp. Phys.",
    volume = "216",
    number = "1-2",
    pages = "29--38",
    year = "2024"
}

@article{BicepKeck:2021ybl,
    author = "Ade, P. A. R. and others",
    collaboration = "Bicep /Keck, Bicep/Keck, BICEP/Keck",
    title = "{Bicep/KeckXV: The Bicep3 Cosmic Microwave Background Polarimeter and the First Three-year Data Set}",
    doi = "10.3847/1538-4357/ac4886",
    journal = "Astrophys. J.",
    volume = "927",
    number = "1",
    pages = "77",
    year = "2022"
}

@article{Thornton:2016wjq,
    author = "Thornton, R. J. and others",
    title = "{The Atacama Cosmology Telescope: The polarization-sensitive ACTPol instrument}",
    doi = "10.3847/1538-4365/227/2/21",
    journal = "Astrophys. J. Suppl.",
    volume = "227",
    number = "2",
    pages = "21",
    year = "2016"
}

@article{McCarrick:2021crh,
    author = "McCarrick, Heather and others",
    title = "{The Simons Observatory Microwave SQUID Multiplexing Detector Module Design}",
    doi = "10.3847/1538-4357/ac2232",
    journal = "Astrophys. J.",
    volume = "922",
    number = "1",
    pages = "38",
    year = "2021"
}

@article{Doriese2016XrayTDM,
author={Doriese, W. B.
and Morgan, K. M.
and Bennett, D. A.
and Denison, E. V.
and Fitzgerald, C. P.
and Fowler, J. W.
and Gard, J. D.
and Hays-Wehle, J. P.
and Hilton, G. C.
and Irwin, K. D.
and Joe, Y. I.
and Mates, J. A. B.
and O'Neil, G. C.
and Reintsema, C. D.
and Robbins, N. O.
and Schmidt, D. R.
and Swetz, D. S.
and Tatsuno, H.
and Vale, L. R.
and Ullom, J. N.},
title={Developments in Time-Division Multiplexing of X-ray Transition-Edge Sensors},
journal={Journal of Low Temperature Physics},
year={2016},
month={Jul},
day={01},
volume={184},
number={1},
pages={389-395},
issn={1573-7357},
doi={10.1007/s10909-015-1373-z},
url={https://doi.org/10.1007/s10909-015-1373-z}
}

@ARTICLE{DurkinAthenaTDM,
  author={Durkin, Malcolm and Adams, Joseph S. and Bandler, Simon R. and Chervenak, James A. and Chaudhuri, Saptarshi and Dawson, Carl S. and Denison, Edward V. and Doriese, William B. and Duff, Shannon M. and Finkbeiner, Fred M. and FitzGerald, Connor T. and Fowler, Joseph W. and Gard, Johnathon D. and Hilton, Gene C. and Irwin, Kent D. and Joe, Young Il and Kelley, Richard L. and Kilbourne, Caroline A. and Miniussi, Antoine R. and Morgan, Kelsey M. and O'Neil, Galen C. and Pappas, Christine G. and Porter, Frederick S. and Reintsema, Carl D. and Rudman, David A. and Sakai, Kazuhiro and Smith, Stephen J. and Stevens, Robert W. and Swetz, Daniel S. and Szypryt, Paul and Ullom, Joel N. and Vale, Leila R. and Wakeham, Nicholas A. and Weber, Joel C. and Young, Betty A.},
  journal={IEEE Transactions on Applied Superconductivity}, 
  title={Demonstration of Athena X-IFU Compatible 40-Row Time-Division-Multiplexed Readout}, 
  year={2019},
  volume={29},
  number={5},
  pages={1-5},
  doi={10.1109/TASC.2019.2904472}}

@article{vanderKuurfMUXAthena,
author = {J. van der Kuur and L. G. Gottardi and H. Akamatsu and B. J. van Leeuwen and R. den Hartog and D. Haas and M. Kiviranta and B. J. Jackson},
title = {{Optimising the multiplex factor of the frequency domain multiplexed readout of the TES-based microcalorimeter imaging array for the X-IFU instrument on the Athena x-ray observatory}},
volume = {9905},
booktitle = {Space Telescopes and Instrumentation 2016: Ultraviolet to Gamma Ray},
editor = {Jan-Willem A. den Herder and Tadayuki Takahashi and Marshall Bautz},
organization = {International Society for Optics and Photonics},
journal = {Proc. SPIE Int. Soc. Opt. Phot.},
publisher = {SPIE},
pages = {99055R},
year = {2016},
doi = {10.1117/12.2232830},
URL = {https://doi.org/10.1117/12.2232830}
}

@Article{Akamatsu2020fMUXAthena,
author={Akamatsu, H.
and Gottardi, L.
and van der Kuur, J.
and de Vries, C. P.
and Bruijn, M. P.
and Chervenak, J. A.
and Kiviranta, M.
and van den Linden, A. J.
and Jackson, B. D.
and Miniussi, A.
and Ravensberg, K.
and Sakai, K.
and Smith, S. J.
and Wakeham, N.},
title={Progress in the Development of Frequency-Domain Multiplexing for the X-ray Integral Field Unit on Board the Athena Mission},
journal={Journal of Low Temperature Physics},
year={2020},
month={May},
day={01},
volume={199},
number={3},
pages={737-744},
issn={1573-7357},
doi={10.1007/s10909-020-02351-3},
url={https://doi.org/10.1007/s10909-020-02351-3}
}

@article{Akamatsu:2021vcs,
    author = "Akamatsu, H. and others",
    title = "{Demonstration of MHz frequency domain multiplexing readout of 37 transition edge sensors for high-resolution x-ray imaging spectrometers}",
    doi = "10.1063/5.0066240",
    journal = "Appl. Phys. Lett.",
    volume = "119",
    number = "18",
    pages = "182601",
    year = "2021"
}

@article{denHartogBBFeedbackSRON,
    author = {den Hartog, Roland and Boersma, D. and Bruijn, M. and Dirks, B. and Gottardi, L. and Hoevers, H. and Hou, R. and Kiviranta, M. and de Korte, P. and van der Kuur, J. and van Leeuwen, B.‐J. and Nieuwenhuizen, A. and Popescu, M.},
    title = {Baseband Feedback for Frequency‐Domain‐Multiplexed Readout of TES X‐ray Detectors},
    journal = {AIP Conference Proceedings},
    volume = {1185},
    number = {1},
    pages = {261-264},
    year = {2009},
    month = {12},
    issn = {0094-243X},
    doi = {10.1063/1.3292328},
    url = {https://doi.org/10.1063/1.3292328},
}

@article{Mates2017uMUX,
    author = {Mates, J. A. B. and Becker, D. T. and Bennett, D. A. and Dober, B. J. and Gard, J. D. and Hays-Wehle, J. P. and Fowler, J. W. and Hilton, G. C. and Reintsema, C. D. and Schmidt, D. R. and Swetz, D. S. and Vale, L. R. and Ullom, J. N.},
    title = {Simultaneous readout of 128 X-ray and gamma-ray transition-edge microcalorimeters using microwave SQUID multiplexing},
    journal = {Applied Physics Letters},
    volume = {111},
    number = {6},
    pages = {062601},
    year = {2017},
    month = {08},
    issn = {0003-6951},
    doi = {10.1063/1.4986222},
    url = {https://doi.org/10.1063/1.4986222},
}

@article{Nakashima2020uMUXSOA,
    author = {Nakashima, Y. and Hirayama, F. and Kohjiro, S. and Yamamori, H. and Nagasawa, S. and Sato, A. and Yamada, S. and Hayakawa, R. and Yamasaki, N. Y. and Mitsuda, K. and Nagayoshi, K. and Akamatsu, H. and Gottardi, L. and Taralli, E. and Bruijn, M. P. and Ridder, M. L. and Gao, J. R. and den Herder, J. W. A.},
    title = {Low-noise microwave SQUID multiplexed readout of 38 x-ray transition-edge sensor microcalorimeters},
    journal = {Applied Physics Letters},
    volume = {117},
    number = {12},
    pages = {122601},
    year = {2020},
    month = {09},
    issn = {0003-6951},
    doi = {10.1063/5.0016333},
    url = {https://doi.org/10.1063/5.0016333},
}

@article{Alpert:2019fit,
    author = "Alpert, B. and others",
    title = "{High-resolution high-speed microwave-multiplexed low temperature microcalorimeters for the HOLMES experiment}",
    doi = "10.1140/epjc/s10052-019-6814-4",
    journal = "Eur. Phys. J. C",
    volume = "79",
    number = "4",
    pages = "304",
    year = "2019"
}

@article{Giachero:2021tli,
    author = "Giachero, A. and others",
    title = "{Progress in the Development of TES Microcalorimeter Detectors Suitable for Neutrino Mass Measurement}",
    doi = "10.1109/TASC.2021.3051104",
    journal = "IEEE Trans. Appl. Supercond.",
    volume = "31",
    number = "5",
    pages = "2100205",
    year = "2021"
}

@article{Ahmine:2023xhg,
    author = "Ahmine, A. and others",
    title = "{Enhanced light signal for the suppression of pile-up events in Mo-based bolometers for the 0$\nu \beta \beta $ decay search.}",
    doi = "10.1140/epjc/s10052-023-11519-6",
    journal = "Eur. Phys. J. C",
    volume = "83",
    number = "5",
    pages = "373",
    year = "2023"
}

@phdthesis{Elleflot:2019lbn,
    author = "Elleflot, Tucker",
    title = "{Measuring the Polarization of the Cosmic Microwave Background with POLARBEAR-1 and Developing the Next-Generation Experiment POLARBEAR-2}",
    school = "UC, San Diego (main)",
    year = "2019"
}

\end{document}